\documentclass[twocolumn,amsmath,amssymb,showpacs]{revtex4}

\usepackage{amsmath} %
\usepackage{graphicx}
\usepackage{amsfonts}
\usepackage{dcolumn}
\usepackage{bm}
\usepackage{color}

\begin{document}
\title{Design principles for biochemical oscillations with limited energy
resources}
\author{Zhiyu Cao}
\author{Huijun Jiang}
\thanks{E-mail: hjjiang3@ustc.edu.cn}
\author{Zhonghuai Hou}
\thanks{E-mail: hzhlj@ustc.edu.cn}
\affiliation{Department of Chemical Physics \textbackslash\& Hefei National Laboratory
for Physical Sciences at Microscales, iChEM, University of Science
and Technology of China, Hefei, Anhui 230026, China}
\date{\today}

\begin{abstract}
As biochemical systems may frequently suffer from limited energy resources
so that internal molecular fluctuation has to be utilized to induce
random rhythm, it is still a great theoretical challenge to understand
the elementary principles for biochemical systems with limited energy
resources to maintain phase accuracy and phase sensitivity. Here,
we address the issue by deriving the energy-accuracy and the sensitivity-accuracy
trade-off relations for a general biochemical model, analytically
and numerically. We find that, biochemical systems consume much lower
energy cost by noise-induced oscillations to keep almost equal efficiency
to maintain precise processes than that by normal oscillations,
elucidating clearly the survival mechanism when energy resources are
limited. Moreover, an optimal system size is predicted where both
the highest sensitivity and accuracy can be reached at the same time,
providing a new strategy for the design of biological networks with
limited energy sources.
\end{abstract}


\maketitle
For a living system to survive and grow, it needs to meet certain
regulatory function and sensory adaptation with energy constantly
injected and dissipated. In particular, for biochemical oscillations
which are crucial in controlling the timing of life processes, such
as cell cycle, circadian clocks, glycolysis, both accurate determination
of the period and sensitive response to external signals are expected
to be ensured\cite{ferrell2011modeling,nakajima2005reconstitution,goldbeter1997biochemical,martiel1987model,buzsaki2004neuronal}. Recently, several experimental findings have implied
that there may be some underlying trade-off relations preventing them
from been reached simultaneously\cite{lan2012energy,endres2009maximum,lan2013cost,sartori2015free,mora2019physical}.
Understanding such relations is then of great importance to uncover
design principles for biochemical oscillations to maintain enhanced
phase accuracy of internal period and phase sensitivity to external
signals. So far, it has been revealed that, for a biochemical oscillation
system with sufficient energy supplies, additional energy exceeding
a critical value can be used to enhance the system's phase accuracy
and phase sensitivity\cite{cao2015free,barato2016cost,fei2018design,hasegawa2019uncertainty}.
However, biochemical oscillation systems in real world may frequently
suffer from limited energy resources\cite{qian2007phosphorylation,guan2020trade,hou2005effects},
so that the critical energy to maintain the oscillatory behavior may
even lack and internal molecular fluctuation has to be utilized to
induce random rhythm. It is then still a great theoretical challenge
to understand the elementary principles for biochemical systems with
limited energy resources to maintain phase accuracy and phase sensitivity.

Here, we address the issue by studying the trade-off relations for
a general biochemical model theoretically, starting from which both
normal oscillation for sufficient energy sources and noise-induced
oscillations for limited energy supplies can be described well in
a unified theoretical framework. By applying the concepts of stochastic
thermodynamics as well as phase reduction method, the energy-accuracy
and the sensitivity-accuracy trade-off relations are finally derived,
which provide general design principles for biochemical oscillations.
Application of these principles shows that the biochemical systems
can keep almost equal efficiency to maintain precise processes at
much lower energy cost by noise-induced oscillations for limited energy
resources than that by normal oscillations for sufficient energy supplies.
Moreover, an optimal system size is found where both the high sensitivity
and high accuracy can be reached at the same time, predicting a new
design strategy for biological networks with limited energy sources.

For a general biochemical system of size $V$ including $N$ well-stirred
species and $M$ reactions $(R_{1},\dots,R_{M})$, its dynamics can
be described by the well-known chemical Langevin equation(CLE)\cite{gillespie2000chemical}
\begin{equation}
\dot{x}_{j}=\sum_{\rho=1}^{M}v_{\rho}^{j}w_{\rho}(\bm{x})+\frac{1}{\sqrt{V}}\sum_{\rho=1}^{M}v_{\rho}^{j}\sqrt{w_{\rho}(\bm{x})}\xi_{\rho}(t),\:j=1,...,N.\label{eq:CLE2}
\end{equation}
where $\bm{x}=(x_{1},\dots,x_{N})^{\text{T}}$ is the concentration
vector, $v_{\rho}^{j}$ is the stochiometric coefficient of $x_{j}$
in reaction $R_{\rho}$, $w_{\rho}(\bm{x})$ is the transition probability,
and $\bm{\xi}(t)$ is independent Gaussian white noises with zero
mean and time correlation $\left\langle \xi_{\rho}(t)\xi_{\rho^{\prime}}(s)\right\rangle =\delta_{\rho\rho^{\prime}}\delta(t-s)$.
The corresponding chemical Fokker-Planck equation (CFPE)\cite{gardiner1985handbook}
is then $\partial_{t}p(\bm{x},\tau)=-\sum_{i}\partial_{x_{i}}[f_{i}(\bm{x})p(\bm{x},\tau)]+\left(1/2\right)\sum_{i,j}V\partial_{x_{i}}\partial_{x_{j}}[G_{ij}(\bm{x})p(\bm{x},\tau)]$
where $p(\bm{x},\tau)$ is the time-varying probability density function,
$G_{ij}(\bm{x})=\sum_{\rho=1}^{M}v_{\rho}^{i}v_{\rho}^{j}w_{\rho}(\bm{x})$,
and the drift $f_{j}(\bm{x})=\sum_{\rho=1}^{M}v_{\rho}^{j}w_{\rho}(\bm{x})$
is the macroscopic rate under thermodynamic limit $V\gg1$. If there
is a Hopf bifurcation for the systems as some control parameters change,
biochemical oscillation will occur. Above the bifurcation, a normal
oscillation will be observed, while near but below the bifurcation
stochastic oscillation can emerge due to the internal noise. A general
theoretical description of the oscillation dynamics including both
the normal one and noise-induced oscillation(NIO)\cite{hou2003internal,hanggi2002stochastic} can be achieved
by the stochastic normal form theory we established before\cite{xiao2007effects,hou2006internal,ma2008coherence},
i.e., the time evolution of oscillation amplitude $r$ and phase $\theta$
is
\begin{equation}
\dot{r}=\alpha r+C_{r}r^{3}+\frac{\varepsilon^{2}}{2Vr}+\frac{\varepsilon}{\sqrt{V}}\eta_{r}(t),\label{eq:dotr}
\end{equation}
\begin{equation}
\dot{\theta}=\omega+C_{i}r^{2}+\frac{\varepsilon}{r\sqrt{V}}\eta_{\theta}(t),\label{eq:dottheta}
\end{equation}
where $C_{r}$ and $C_{i}$ are negative constants, $\varepsilon^{2}$
is the averaged noise intensity, $\eta_{r}(t)$ and $\eta_{\theta}(t)$
are the averaged independent Gaussian white noises with unit variances
(see details in the supplemental information, SI). Specially, $\alpha$
can be related to the energy resources \cite{guan2020trade} which
determines the oscillatory behaviors of the systems. For large enough
$V$, normal oscillations with the amplitude $r_{m}^{2}=-\frac{\alpha}{C_{r}}+\frac{\varepsilon^{2}}{2\alpha}V^{-1}+o(V^{-2})$
can be observed when $\alpha>0$. When $\alpha<0$, the energy resources
are not enough to support normal oscillations. However, there is still
a nonzero amplitude solution $r_{m}^{2}=\frac{\varepsilon^{2}}{2\alpha}V^{-1}+o(V^{-2})$,
indicating that the internal noise could be utilized to induce stochastic
rhythms to maintain the system's function\cite{ko2010emergence,hou2005effects}
in such a situation.

Energy-accuracy trade-off relation describes the constraint that how
accurate a biochemical oscillator can be with given energy sources.
As the time translation symmetry of the biochemical system is inherently
broken, the phase of the oscillation exhibits a diffusive behavior.
The coefficient of phase diffusion $D_{\theta}$ can be used to measure
the accuracy of the oscillators. By Eq. (\ref{eq:dotr}) and (\ref{eq:dottheta}),
one can derive that the steady-state probability distribution reads
$p_{s}(r)=C_{0}r\exp\left[\left(\alpha r^{2}+\frac{1}{2}C_{r}r^{4}\right)/\left(\varepsilon^{2}/V\right)\right]$
with $C_{0}$ a normalization constant. Notice that, $p_{s}(r)$ shows
a maximum at $r=r_{m}$, indicating that there is an attractor of
limit cycle and the system will fluctuate around it due to internal
noise. Then, one can calculate the mean and variance of $\theta$,
i.e., $\left\langle \theta(t)\right\rangle =\left(\omega+C_{i}r_{m}^{2}\right)t\equiv\omega_{s}t$
and $\left\langle \theta(t)^{2}\right\rangle -\left\langle \theta(t)\right\rangle ^{2}\simeq\varepsilon^{2}t/\left(Vr_{m}^{2}\right)$,
where $\omega_{s}=\omega+C_{i}r_{m}^{2}$ is the effective phase angular
velocity of the attractor. Thus, the diffusion constant of the phase
fluctuation is given by
\begin{equation}
D_{\theta}=\varepsilon^{2}/\left(Vr_{m}^{2}\right).\label{eq:Dtheta1}
\end{equation}
Besides, the explicit expression for energy dissipation in one cycle
is $\text{\ensuremath{\Delta W}}=\dot{S}_{tot}T_{cyc}$ with $\dot{S}_{tot}$
the total entropy production rate and $T_{cyc}=2\pi/\omega_{s}$ the
period of the oscillation. At last, by applying the concepts of stochastic
thermodynamics\cite{seifert2005entropy}, the total entropy production
rate is found to be $\dot{S}_{tot}=\left(k_{b}+k_{a}Vr_{m}^{2}\right)/(2\pi),$
where $k_{a}$, $k_{b}$ are system dependent parameters (see Supplemental
information for details). Now, we arrive at the first main result
of this paper, namely, the energy-accuracy trade-off relation (see
SI for details)
\begin{equation}
D_{\theta}=\begin{cases}
\frac{k_{a}\varepsilon^{2}}{\omega}\left(\Delta W+\frac{k_{b}C_{r}}{C_{i}\alpha-C_{r}\omega}\right)^{-1}+D_{\theta,-}^{s} & \text{\ensuremath{\alpha<0}}\\
\frac{k_{a}\varepsilon^{2}}{\omega}\left(\Delta W-\frac{k_{b}}{\omega V}\right)^{-1}+D_{\theta,+}^{s} & \alpha>0
\end{cases}.\label{eq:DthetaDW}
\end{equation}
Here, Eq.(\ref{eq:DthetaDW}) provides a general design principle
to quantitatively describe the balance between energy cost and phase
accuracy for not only normal oscillations ($\alpha>0$) but also NIOs
($\alpha<0$).

Several conclusions can be obtained. Firstly, Eq.(\ref{eq:DthetaDW})
can recover to the reported one for normal oscillations\cite{cao2015free}
$D_{\theta}=W_{0}/(\Delta W-W_{c})$ with $W_{0}=k_{a}\varepsilon^{2}/\omega$,
$W_{c}=-k_{b}C_{r}/\left(C_{i}\alpha-C_{r}\omega\right)$ and an additional
$D_{\theta}^{s}$ (for $\alpha>0$, $D_{\theta,-}^{s}=-C_{i}\varepsilon^{2}/(\omega V)$;
for $\alpha<0$, $D_{\theta,+}^{s}=-2\alpha$). Secondly, Eq.(\ref{eq:DthetaDW})
shows that phase diffusion can be suppressed by increasing thermodynamic
cost $\Delta W$, while there is always a minimal phase diffusion
constant $D_{\theta}^{s}$ that cannot be completely eliminated even
for infinite cost.

Thirdly, as $D_{\theta}=\varepsilon^{2}/\left(Vr_{m}^{2}\right)$,
the scaling law for the energy cost $\Delta W$ of the system size
$V$ can be derived from Eq.(\ref{eq:DthetaDW}) as
\begin{equation}
\Delta W\sim V^{\nu},\begin{cases}
\nu=0 & \alpha<0\\
\nu=\frac{1}{2} & \alpha=0\\
\nu=1 & \alpha>0
\end{cases},\label{eq:sl2}
\end{equation}
i.e., for normal oscillations, $\Delta W$ increases linearly as $V$
increases, while for NIOs $\Delta W$ is independent on the system
size $V$. Finally, a transport efficiency $\eta_{T}$ \cite{dechant2018current}
quantitatively determining the efficiency of consuming energy to maintain
precise processes for biochemical networks can be derived from the
energy-accuracy trade-off relation Eq.(\ref{eq:DthetaDW}) via the
thermodynamic uncertainty relation (see SI
for details)\cite{barato2015thermodynamic,pietzonka2016universal,horowitz2019thermodynamic}
\begin{equation}
\eta_{T}=\frac{\text{\ensuremath{\langle \dot{\theta}\rangle^{2} }}}{D_{\theta} \dot{S}_{tot}}.\label{eq:etaT}
\end{equation}
More importantly, according to Eq.(\ref{eq:sl2}) and (\ref{eq:etaT}),
it will be found that the biochemical systems need significantly less
energy cost to maintain accuracy for NIOs than for normal ones, which
will be elucidated more clearly in numerical simulations.

As biochemical oscillations with high sensitivity
are vulnerable to external perturbation and fluctuation,
accuracy and sensitivity can be treated as  two trade-off properties\cite{hasegawa2014optimal,fei2018design,hasegawa2014circadian,hasegawa2019uncertainty}.
To achieve the sensitivity-accuracy relation, we now try
to figure out the sensitivity $\chi$ as follows. The deterministic
evolution equation of Eq.(\ref{eq:CLE2}) can be expressed as $\dot{\phi}=\nabla_{\bm{x}}\phi\cdot\bm{f}(\bm{x}).$
In the presence of an external signal $\bm{\beta}(t)$, the deterministic term
obeys $f_{k}(\bm{x})=f(\bm{x})+k\bm{\beta}(t),$ where $k$ is a parameter
to be perturbed. Then, the phase response curve function (PRC) $Z_{k}(\phi)=\nabla_{\bm{x}}\phi\cdot\bm{\beta}(t)$
characterizing the ability of the biochemical circuits to response
to external signals is obtained by comparing the phase shift after
delivering a perturbation at a given duration of time\cite{buhr2010temperature,saunders1994light,johnson1992phase}. By similar
definition of phase as in Ref.\cite{kuramoto2003chemical,goldobin2010dynamics} and the stochastic normal form theory, the sensitivity is then defined as the normalized value of key signal-independent factor $\nabla_{\bm{x}}\phi$ in PRC\cite{fei2018design}
at $r=r_{m}$ (see SI for details),
\begin{equation}
\chi_{m}=\chi(r_{m})=-\frac{C_{i}r_{m}^{2}}{\sqrt{(\alpha^{2}-2C_{r}\varepsilon^{2}/V)}}\label{eq:chirm}.
\end{equation}
Based on Eq.(\ref{eq:chirm}), we can derive the
second main result of this paper
\begin{equation}
2\text{\ensuremath{\log\chi_{m}^{*}}}= C_{0}+\log D_{\theta}\label{eq:tor2}
\end{equation}
with $C_{0}=\log\left(2\pi C_{i}^{2}r_{m}^{4}/\left[\varepsilon^{2}k_{a}(\alpha^{2}-2C_{r}\varepsilon^{2}/V)\right]\right)$ and $\chi_{m}^{*}=\chi_{m}/\sqrt{\dot{S}_{tot}}$.
Eq.(\ref{eq:tor2}) shows that an increase in phase accuracy $D_{\theta}^{-1}$
cannot be accompanied by an increase in the normalized phase sensitivity
$\chi_{m}^{*}$, which can be treated as the sensitivity-accuracy
trade-off relation under fixed energy condition (see SI for detailed analysis of Eq.(\ref{eq:tor2})). Such trade-off
relationship always holds when system parameters change, providing
another design principle for biochemical systems to measure the balance
between the sensory to external signals and the regulatory of the
internal oscillation. More interestingly, we can also obtain a scaling law for phase sensitivity
$\chi_{m}$ of the system size $V$ as
\begin{equation}
\chi_{m}\sim V^{\kappa},\begin{cases}
\text{\ensuremath{\kappa}}=-1 & \alpha<0\\
\text{\ensuremath{\kappa}}=-\frac{1}{2} & \alpha=0\\
\kappa=0 & \alpha>0
\end{cases},\label{eq:sl3}
\end{equation}

In addition, a dynamic efficiency can be defined to quantitatively describe
the ability of biochemical systems to maintain the sensory adaptation
for external signals and the regulatory of the internal function at
the same time
\begin{equation}
\eta_{S}=\frac{\chi_{\text{\ensuremath{m}}}^{2}}{D_{\theta}}=\frac{C_{i}^{2}Vr_{m}^{6}}{\varepsilon^{2}(\alpha^{2}-2C_{r}\varepsilon^{2}/V)}.\label{eq:etaS}
\end{equation}

The obtained dynamic efficiency can further be related to the information inequality $v_{k,\phi}^{2}/D_{\theta}\leq\dot{\mathbb{D}}_{PE}$
where $v_{k,\phi}$ is the change rate of the current difference $\left\langle \phi\right\rangle _{k}-\left\langle \phi\right\rangle $ (here $\left\langle \phi\right\rangle $ is the mean of phase for
the original dynamics, $\left\langle \phi\right\rangle _{k}$ is
the mean for the dynamics perturbed by the signal), and $\mathbb{D}_{PE}$
is the Pearson divergence between the original dynamics and the perturbed
dynamics which show similar evolutionary behavior to the total entropy production\cite{hasegawa2019uncertainty}. The quantity $v_{k,\phi}$
is proportional to the system sensitivity, $\dot{\mathbb{D}}_{PE}$ determines the upper bound of the dynamic efficiency. Moreover, as the total entropy production rate hardly changes with system size in NIOs, we find that dynamic efficiency can be approached to upper bound by adjusting the size parameter $V$ of biochemical oscillation systems. Therefore, one can design a biochemical system to enhance the sensitivity and
reduce the fluctuation simultaneously by changing its internal properties
to maximize $\eta_{S}$. Interestingly, an optimal system size $V_{opt}$
can be achieved by setting $\partial\left(\eta_{S}\right)/\partial V=0$,
where biochemical systems with limited energy sources can reach their
best performance of both high sensitivity and high accuracy.

Now, we apply the above analytical results to a well-known biochemical
oscillation system, the Brusselator model involving two distinct biochemical
species $X_{1}$,$X_{2}$, four reaction channels,
$$
A\to X_{1},\qquad B+X_{1}\to X_{2},
$$
$$
X_{1}\to C,\qquad2X_{1}+X_{2}\to3X_{1}.
$$
Here $\bm{w}=(A,BX_{1},X_{1},X_{1}^{2}X_{2})$ represent the corresponding transition rates.
In the thermodynamic limit where the internal noise terms can be ignored,
a supercritical Hopf bifurcation occurs for $\alpha=(B-B_{c})/2$
with $B_{c}=A^{2}+1$. We numerically simulate Eq.(\ref{eq:CLE2})
by Euler methods with a time step $10^{-4}$. After long enough transition
time, $10^{5}$ trajectories are used to calculate the energy cost $\Delta W$.

\begin{figure}
\includegraphics[scale=0.45]{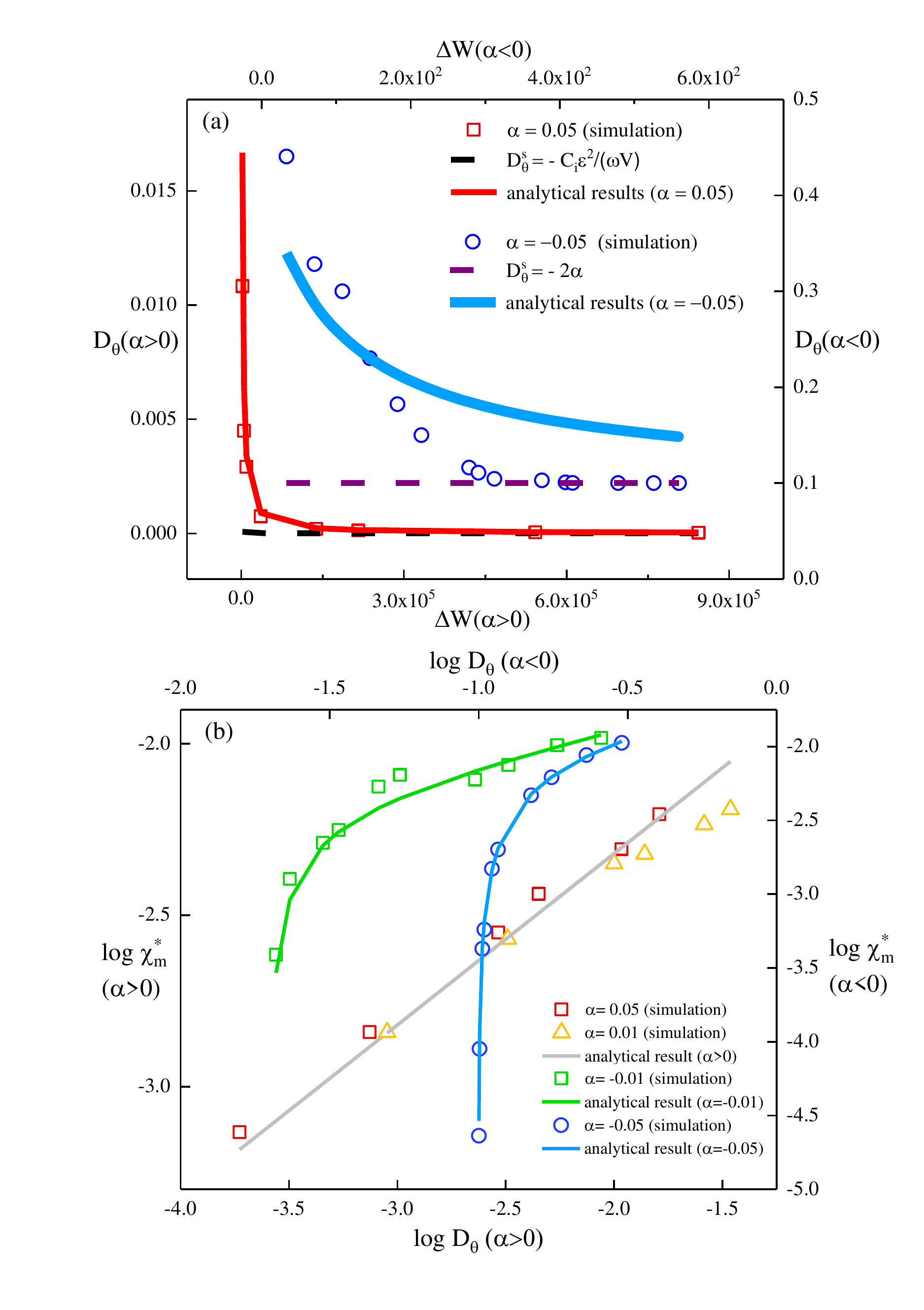}
\caption{\textbf{Trade-off relations for biochemical oscillations with sufficient
($\alpha>0$) or limited ($\alpha<0$) energy sources. }(a) The simulated
energy-accuracy trade-off relation for \textbf{$\alpha>0$} and \textbf{$\alpha<0$}
shows that the phase fluctuation constant $D_{\theta}$ decreases
as the free energy cost $\Delta W$ increases, which agrees well with
analytical curves fitted according to Eq.(\ref{eq:DthetaDW}). Besides,
there is always a minimal phase diffusion constant $D_{\theta}^{s}$
(the black dash line for $\alpha>0$ and the purple dash line for $\alpha<0$
in (a)) cannot be completely eliminated even for infinite cost. (b)
The simulated sensitivity-accuracy trade-off relation for \textbf{$\alpha>0$}
and \textbf{$\alpha<0$} shows that phase sensitivity $\chi_{m}$
also increases as the phase diffusion constant $D_{\theta}$ increases.}
\label{fig:plot-1}
\end{figure}

Both the trade-off relations Eq.(\ref{eq:DthetaDW}) and (\ref{eq:tor2})
are shown in Fig.\ref{fig:plot-1}. It can be observed that, the phase
diffusion constant $D_{\theta}$ is inversely proportional to the
energy cost $\Delta W$, confirming that the accuracy-energy trade-off
relation holds for both NIOs and normal oscillations (Fig.\ref{fig:plot-1}(a)).
Similarly, the normalized sensitivity $\chi_{m}^{*}$ also increases
as $D_{\theta}$ increases, which verifies the sensitivity-accuracy
trade-off relation for both normal oscillations and NIOs (Fig.\ref{fig:plot-1}(b)).

\begin{figure}
\begin{centering}
\includegraphics[scale=0.45]{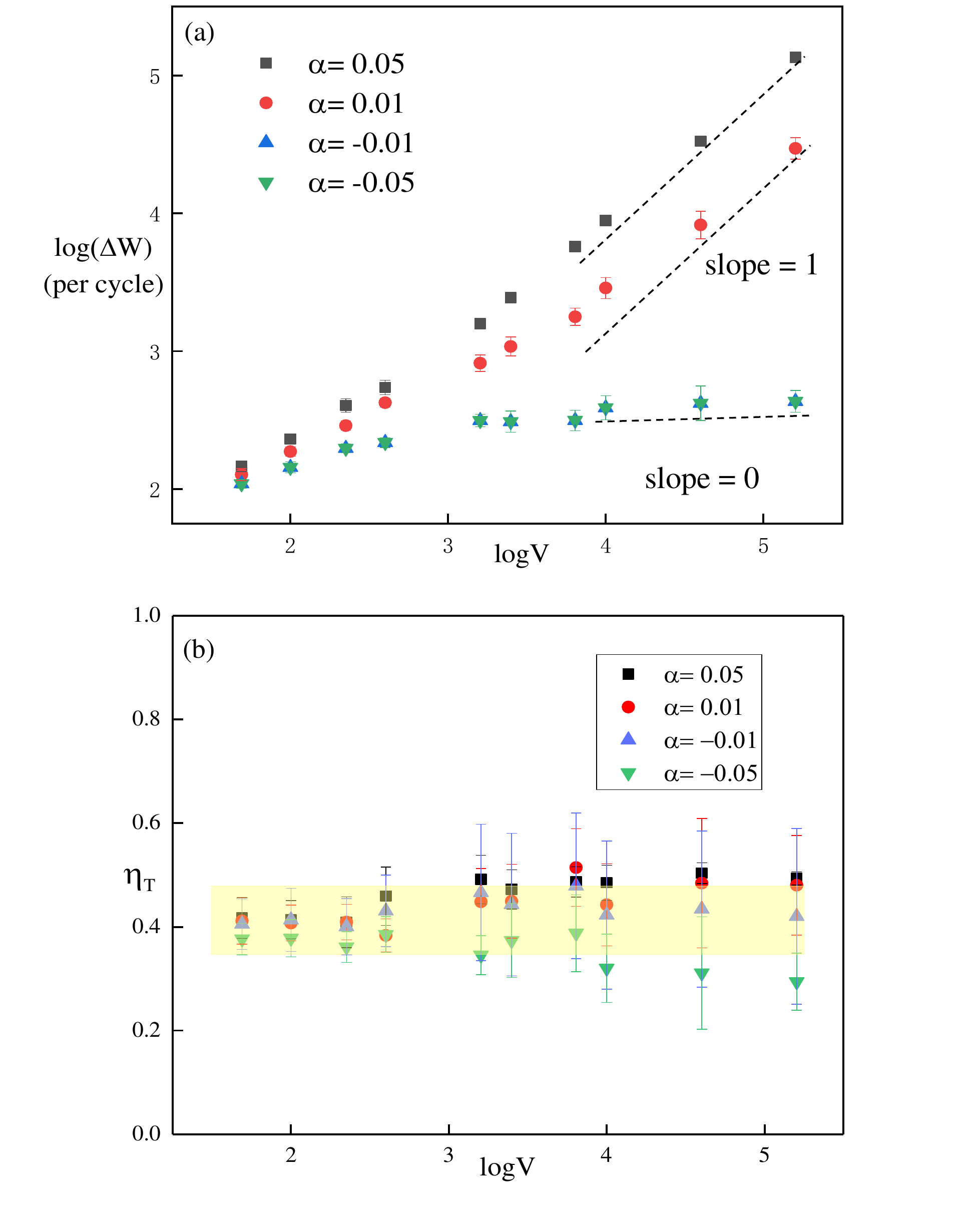}
\par\end{centering}
\caption{\textbf{Biochemical systems with limited energy sources ($\text{\ensuremath{\alpha<0}}$)
need significantly less energy to maintain accuracy than that with
sufficient sources ($\text{\ensuremath{\alpha>0}}$). }(a) Scaling
law between energy cost per cycle $\Delta W$ and system size $V$.
(b) Transport efficiency $\text{\ensuremath{\eta_{T}}}$ for different
system size $V$. Error bars in (a) and (b) represent standard deviations.
For fixed $V$, $\text{\ensuremath{\eta_{T}}}$ is robust against
changes in $\alpha$ and $V$, while $\Delta W$ is much smaller for
$\alpha<0$ than that for $\alpha>0$.}
\label{fig:plot-2}
\end{figure}

Dependence of the energy cost $\Delta W$ and the transport efficiency
$\eta_{T}$ on the system size $V$ is shown in Fig.\ref{fig:plot-2}(a)
and (b), respectively. Simulated $\Delta W$ increases proportionally
as $V$ increases for normal oscillations and is nearly unchanged
for NIOs, which is in good consistence with the scaling law (\ref{eq:sl2})
(Fig.\ref{fig:plot-2}(a)). Additionally, as $\eta_{T}$ is almost
equal for different types of oscillations (Fig.\ref{fig:plot-2}(b))
and $\Delta W$ for NIOs is always smaller than that for normal oscillations
for fixed $V$, it thus leads to a quite interesting conclusion that,
for NIOs, the system can keep almost the same efficiency to maintain
precise processes at much lower energy cost, elucidating clearly the
advantage of noise-induced oscillations in limited energy supplies.

\begin{figure}
\includegraphics[scale=0.7]{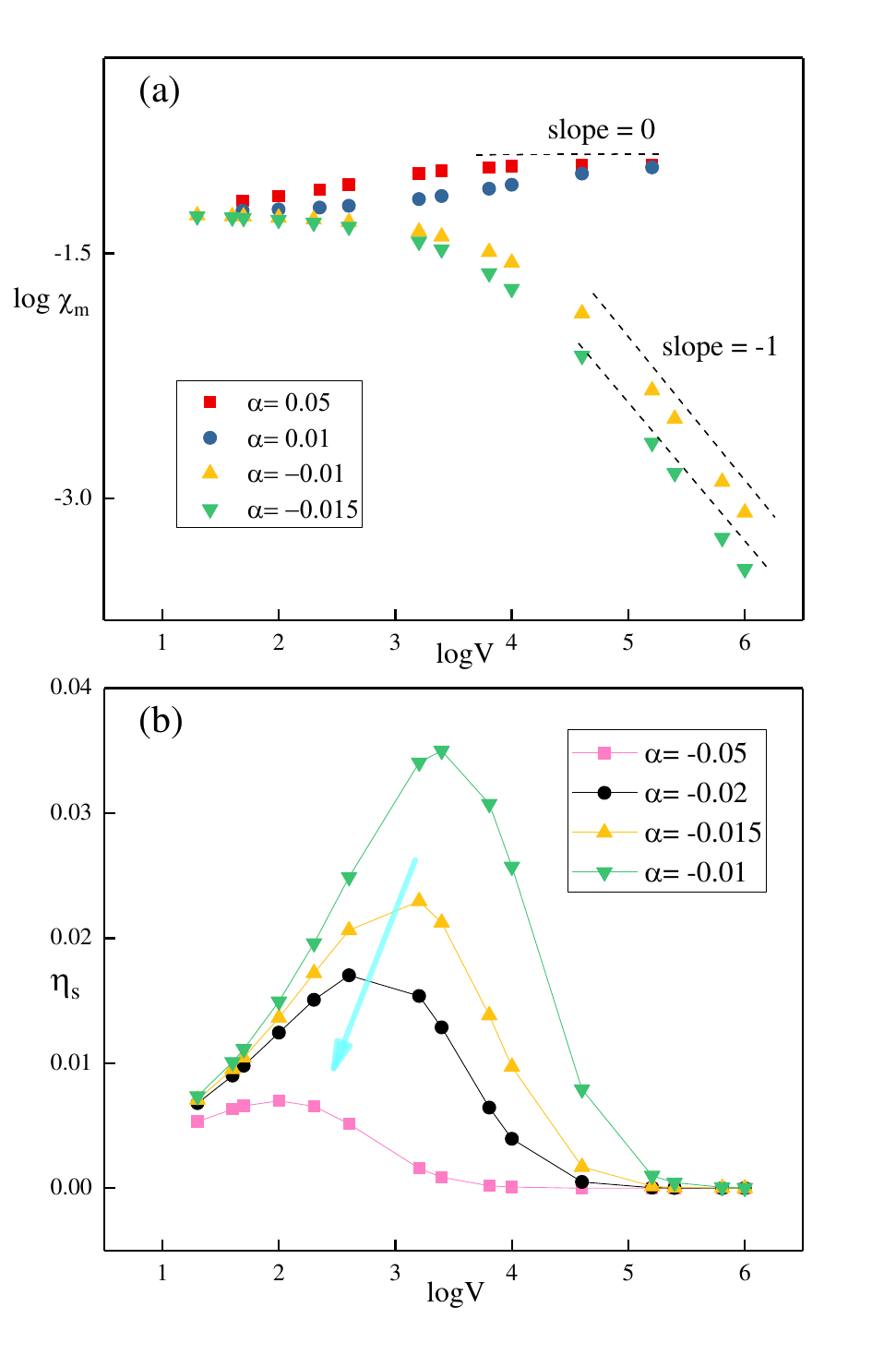}
\caption{\textbf{A new strategy for the design of biological networks in limited
energy resources to achieve both high accuracy and sensitivity. }(a)
Scaling law between the energy cost per cycle $\chi_{m}$ and the
system size $V$ for sufficient ($\alpha>0$) or limited ($\alpha<0$)
energy sources. (b) Dependence of the dynamic efficiency $\eta_{S}$
on the system size $V$. Remarkably, the dynamic efficiency $\eta_{S}$
for limited energy resources shows a maximum for an optimal system
size $V_{opt}$ where $\chi_{m}$ still changes little.}
\label{fig:plot-4}
\end{figure}

As shown in Fig.\ref{fig:plot-4}(a), the phase sensitivity $\chi_{m}$
changes little with increasing system size $V$ for normal oscillations
and is inversely proportional to $V$ for NIOs, agree with the scaling
law Eq.(\ref{eq:sl3}) very well. Remarkably, the dynamic efficiency
$\eta_{S}$ for NIOs in Fig.\ref{fig:plot-4}(b) shows a maximum for
an optimal system size $V_{opt}$ where $\chi_{m}$ still changes
little. This finding predicts a new strategy for the design of biological
networks in limited energy resources to achieve both high accuracy
and sensitivity, simultaneously, which is absent for systems with
sufficient energy supplies (see SI for details).

In conclusion, energy-accuracy and sensitivity-accuracy trade-off
relations have been revealed for a general biochemical system by applying
the framework of stochastic thermodynamics as well as phase reduction
method. According to these relations, it was found that biochemical
systems may maintain their necessary regulatory function via noise-induced
oscillation to save the energy cost when energy resources limited.
More interestingly, an optimal system size for systems to achieve
both high accuracy and sensitivity has also been derived by the trade-off
relations, predicting a new strategy for the design of biological
networks with limited energy sources. As our findings are of important
relevance to many rhythmic processes in biochemical systems, and can
be extended to other realistic systems straightforwardly, it is our
hope that the reported principles will certainly enhance our ability
in designing new biochemical systems for practical applications.

\begin{acknowledgments}
This work is supported by MOST(2016YFA0400904, 2018YFA0208702), NSFC
(21973085, 21833007, 21790350, 21673212, 21521001, 21473165), the
Fundamental Research Funds for the Central Universities (WK2340000074),
and Anhui Initiative in Quantum Information Technologies (AHY090200).
\end{acknowledgments}

\appendix


\end{document}